\documentclass[fleqn,usenatbib]{mnras}
\usepackage{newtxtext,newtxmath}

\usepackage[T1]{fontenc}

\DeclareRobustCommand{\VAN}[3]{#2}
\let\VANthebibliography\thebibliography
\def\thebibliography{\DeclareRobustCommand{\VAN}[3]{##3}\VANthebibliography}

\usepackage{graphicx}	
\usepackage{amsmath}	
\usepackage{url}
\usepackage{threeparttable}
\usepackage{float}
\usepackage{placeins}
\usepackage{orcidlink}

\newcommand{\ra}[3]{$#1^{\mathrm{h}}\,#2^{\mathrm{m}}\,#3^{\mathrm{s}}$}
\newcommand{\dec}[3]{$#1^{\mathrm{d}}\,#2^{\mathrm{m}}\,#3^{\mathrm{s}}$}
\newcommand{\radec}[6]{\ra{#1}{#2}{#3}\quad\dec{#4}{#5}{#6}}

\title[New velocities for local galaxy candidates]{New optical velocities for local galaxy candidates}

\author[M. I. Chazov et al.]{
Maksim I. Chazov,$^{1}$\orcidlink{0009-0004-4126-8924}\thanks{E-mail: chazovmi@gmail.com (MIC)}
Igor D. Karachentsev,$^{1}$\orcidlink{0000-0003-0307-4366}
Serafim S. Kaisin$^{1}$
\\
$^{1}$Special Astrophysical Observatory,  Nizhnij Arkhyz,  Zelenchukskiy region,  Karachai-Cherkessian Republic, 369167, Russia\\
}

\date{Accepted XXX. Received YYY; in original form ZZZ}

\pubyear{\the\year{}}

\begin{document}
\label{firstpage}
\pagerange{\pageref{firstpage}--\pageref{lastpage}}
\maketitle

\begin{abstract}

  We used the 6-meter BTA telescope to determine radial velocities for 46 galaxies, previously considered to be nearby objects.
  Twelve of them have kinematic distances within 13\,Mpc and are new probable companions to the bright Local Volume galaxies: NGC\,1068, NGC\,2903, NGC\,4517, NGC\,4565, NGC\,4826.
  We also found a new tiny satellite of the nearby (10.38\,Mpc) SMC-like dwarf galaxy DDO\,46.
  Two isolated dIrr galaxies: dw\,1408-0802 and Dw\,1552-1027 are probably located inside the Local Void.
  The spectrum of the latter shows a rich set of emission lines.
  Using the radial velocities and projected separations of 13 NGC\,1068 satellites,
  we estimated the total mass of the group to be $M= (2.6\pm1.0)\times10^{12}\, M_\odot$.

\end{abstract}




\section{Introduction}

Radial velocity anchors a galaxy's position in phase space and underpins dynamical studies from satellite systems to the local peculiar-velocity field.
Major optical and radio surveys --- SDSS~\citep{Abazajian2009}, HIPASS~\citep{Koribalski2004}, ALFALFA~\citep{Haynes2011}, FAST/FASHI~\citep{Zhang2024} --- have progressively expanded the redshift catalogue,
and the latest multiplexed spectroscopy has pushed the total yield to tens of millions~\citep{DESICollab2025}.
Yet nearby dwarf galaxies, especially low-surface-brightness systems, remain kinematically incomplete:
selection biases in optical spectroscopy and practical limitations of blind 21-cm surveys at very low heliocentric velocities leave a persistent gap.

The Local Volume (LV; $D\leq11$--12\,Mpc) is the regime where this incompleteness matters most.
The UNGC/LVGDB\footnote{\url{http://www.sao.ru/lv/lvgdb}}~\citep{Karachentsev2013} currently lists more than ${\sim}1500$ LV galaxies,
yet radial velocities are still missing for a significant fraction.
The DESI Legacy Imaging Surveys~\citep{Dey2019} have accelerated the discovery of new dwarf candidates \citep[e.g.][]{Karachentsev2024, Karachentsev2025mnras};
targeted follow-up is beginning to close the kinematic gap:
\citet{Karachentsev2025mnras} obtained optical long-slit velocities for 40 candidates and identified multiple new LV companions,
while \citet{Nazarova2025} measured HI velocities for 105 nearby dwarfs with the GBT (77 detections).
Where velocities are available, they provide the basis for preliminary kinematic distances via the Numerical Action Method (NAM)~\citep{Shaya2017, Kourkchi2020}.

Here we present new BTA long-slit radial velocities for 46 LV dwarf candidates selected from the DESI Legacy Imaging Survey,
with the aim of improving velocity completeness, refining group membership, and providing NAM distance estimates.
The observations and reduction are described in Section~\ref{sec:obs}; results are given in Section~\ref{sec:results};
Section~\ref{sec:discussion} discusses individual objects and the NGC\,1068 group dynamics; Section~\ref{sec:conclusions} summarises our conclusions.

\section{Observations with the 6-meter BTA telescope}
\label{sec:obs}

We performed long-slit observations at the 6-m telescope of the Special Astrophysical Observatory of
the Russian Academy of Sciences with the multi-mode focal reducers SCORPIO-1~\citep{Afanasjev2005} and
SCORPIO-2~\citep{Afanasjev2011}.
The VPHG1200R grating was used with SCORPIO-1, providing a spectral resolution of 6\,\AA{} over 5700--7500\,\AA{},
while the VPHG1200@540 grating was used with SCORPIO-2, providing a spectral resolution of 5.2\,\AA{} over 3650--7300\,\AA{}.
The slit width and length were $(1.0$–$1.2)^{\prime\prime}\times5^{\prime}$, respectively.
The observations were carried out in late 2024, throughout 2025 and in early 2026 under typical seeing of $1$–$2^{\prime\prime}$,
with exposure times of 300–2700\,s.

We performed data reduction using standard procedures implemented in an IDL-based pipeline, as described in detail by~\citet{Egorov2018}.
Line-of-sight velocities along the slit were measured by fitting Gaussian profiles to the H$\alpha$ and other emission lines, when the S/N ratio was sufficiently high.
The barycentric velocity correction\footnote{\url{https://docs.astropy.org/en/stable/coordinates/velocities.html}} was applied using the Astropy package~\citep{AstropyColl2013}.
We computed the Local Group correction assuming a Solar apex with $l = 93^{\circ}$, $b = 4^{\circ}$, and $v_{\odot} = 316\,\mathrm{km\,s^{-1}}$~\citep{Karachentsev1996}.
The typical accuracy of the radial velocity measurements was ${\sim}10\,\mathrm{km\,s^{-1}}$ for galaxies with strong emission lines.
Since our observations were carried out episodically over 2024--2026, a significant overlap with objects from the DESI spectroscopic survey~\citep{DESICollab2025} and \citet{Nazarova2025} was found a posteriori.

\section{Results}
\label{sec:results}

\begin{figure*}
     \centering
     \includegraphics[width=\textwidth]{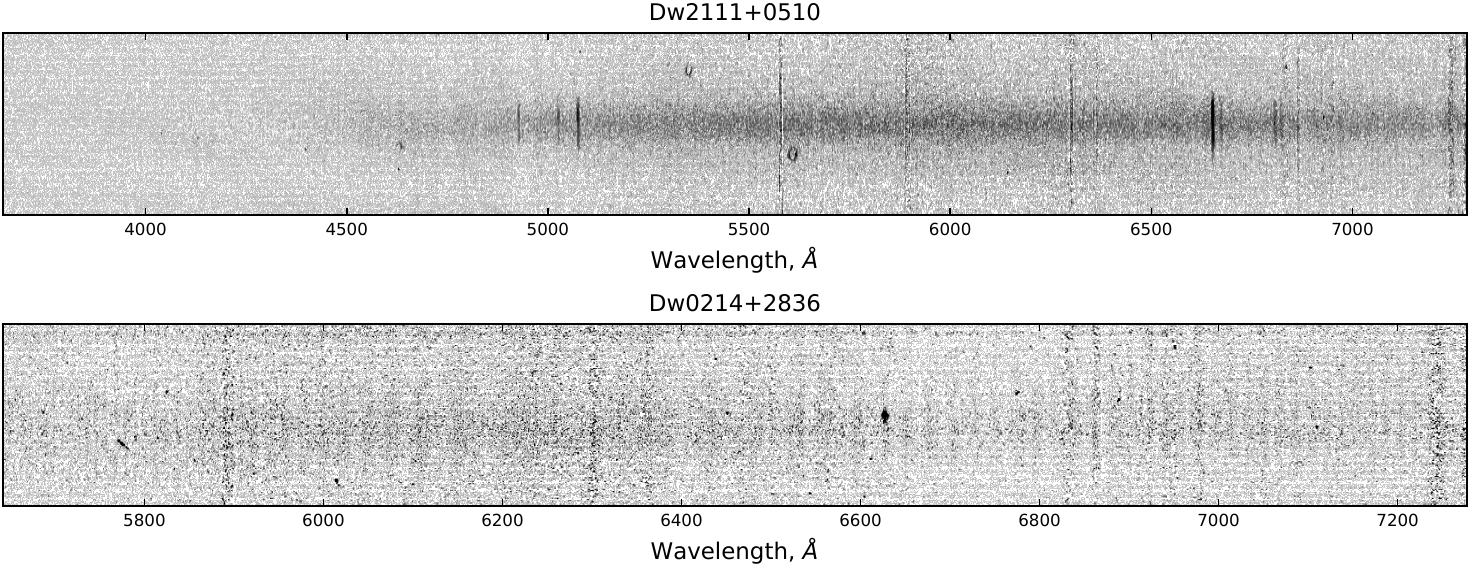}
     \caption{Representative spectrum of a galaxy with strong emission lines obtained with SCORPIO-2 (Dw\,2111+0510) and a galaxy with only weak H$\alpha$ obtained with SCORPIO-1 (Dw\,0214+2836).}
     \label{fig:example_spectrums}
 \end{figure*}

We observed a total of 62 galaxies. Strong emission lines (at least H$\alpha$) are present in 34 objects,
12 additional galaxies show only weak H$\alpha$ emission with very low signal-to-noise ratios,
and 16 galaxies do not exhibit detectable emission lines.
Representative spectra for both cameras are shown in Fig.~\ref{fig:example_spectrums}.
The results of the measured radial velocities are presented in Tables~\ref{tab:lv_galaxies} and \ref{tab:survey_galaxies}.

The kinematic distances in column (7) were derived using the NAM Distance–Velocity Calculator~\citep{Kourkchi2020},
which computes expectation distances from radial velocities based on a smoothed local velocity field within 38 Mpc.
The underlying model~\citep{Shaya2017} reconstructs physically consistent,
fully nonlinear gravitational orbits for 1382 galaxies and groups,
constrained by Cosmicflows survey data,
thereby correcting for peculiar motions taking into account the local flow towards the Virgo cluster and the Local Void expansion that would otherwise bias a simple Hubble-law distance conversion.
The distances measured in the current work include a systematic correction of $\Delta D=+0.3$~Mpc,
derived from a comparison of NAM and TRGB distances for 430 LV galaxies~\citep{Karachentsev2025ap}.
The typical random uncertainty of individual NAM distances within the LV is 1.6~Mpc,
applicable to galaxies lying outside the $\sim25^{\circ}$ exclusion zone around the Virgo cluster.

 \section{Discussion}
\label{sec:discussion}

 Among the observed dwarf galaxies with measured radial velocities, 12 objects have NAM distances less
 than 13\,Mpc. Two and five galaxies in our sample are identified as members of the NGC\,1068
and NGC\,4151 groups, respectively.

\subsection{Description of individual galaxies}

    {\em dw\,0139+1433}  This is a blue irregular dwarf, a peripheral member of the NGC\,628 group.
According to~\citet{Carlsten2022}, its distance is 10.82 Mpc estimated via surface brightness
fluctuations.
Its optical velocity obtained by us agrees well with a recently obtained HI velocity~\citep{Nazarova2025}.

  {\em MAGE\,0148+2830, MAGE\,0202+2717} Two blue compact galaxies from the list by
\citet{Hunter2025}, where the dwarfs were erroneously assumed to be satellites of NGC\,672 and
NGC\,784, respectively.

  {\em Dw\,0214+2836}  A probable satellite of the spiral Sb galaxy NGC\,865, with a radial velocity of $V_h= 2907 \pm 1$\,km\,s$^{-1}$. At a distance of $36.5$~Mpc, the projected separation, $28\arcmin$, is $297$~kpc.

  {\em SMDG\,0223-0203}  An isolated irregular dwarf from the catalog by~\citet{Zaritsky2021}.

  {\em Dw\,0229-0318, Dw\,0232-0256} Background dwarf irregular galaxies near the NGC\,1068 group.

  {\em FASHI\,0237+3855} This dIrr dwarf has a radial velocity difference of $\Delta V_h = 100$\,km\,s$^{-1}$ relative to the edge-on Sb galaxy NGC\,891, which has a radial velocity of $V_h = 528 \pm 3$\,km\,s$^{-1}$ and a distance of $9.95$~Mpc via TRGB. Their projected separation is $4.5$~deg, or $780$~kpc. The escape velocity for a Milky-Way-like galaxy at a distance of $800$~kpc is about $100$\,km\,s$^{-1}$, which suggests that this dwarf could be associated with it.

  {\em Dw\,0242+0133, Dw\,0243-0108} Two new dIrr members of the NGC\,1068 group.

  {\em PGC\,1166738} An isolated irregular dwarf.

  {\em dw\,0445+0344} A field dwarf with sbf-distance of 5.46$\pm$0.5 Mpc~\citep{Carlsten2026}.

  {\em SMDG\,0740+4032, dw\,0741+4005} Two new dIrr companions of the dwarf galaxy DDO\,46, having a radial velocity of $361 \pm 1$\,km\,s$^{-1}$ and the TRGB distance of $10.38$~Mpc. With respect to DDO\,46, they have radial velocity differences of $+2 \pm 10$\,km\,s$^{-1}$ and $-74 \pm 16$\,km\,s$^{-1}$, and projected separations of $89$~kpc and $7$~kpc, respectively.

  {\em dw\,0936+2135} A new spheroidal (dSph) satellite of the bright spiral galaxy NGC\,2903, which has a TRGB distance of $9.15$~Mpc and radial velocity of $V_h= 550 \pm 1$\,km\,s$^{-1}$. Their radial velocity difference is $+94 \pm 38$\,km\,s$^{-1}$ and a projected separation of $158$~kpc.
  
  {\em dw\,1008+7038b}  Its high NAM distance is inconsistent with the sbf distance of 10.63 Mpc from~\citet{Li2025}.

  {\em [KK98]\,108}   A low-surface-brightness dwarf galaxy in the field.

  {\em MAGE\,1216+6906} A dIrr galaxy in the field behind the bright spiral NGC\,4236.

  {\em dw\,1237+2602} A dIrr satellite of the spiral galaxy NGC\,4565.
   According to~\citet{Carlsten2022}, its distance is estimated at 10.97 Mpc via surface brightness
   fluctuations.

  {\em KDG\,171} A transition-type (Tr) dwarf at an sbf distance of 8.34 Mpc~\citep{Carlsten2023}.
It is a satellite of the spiral galaxy NGC\,4517 ($D = 8.36$\,Mpc via TRGB, $V_h = 1134$\,km\,s$^{-1}$). The high
radial velocities of both galaxies are caused by their infall towards the Virgo cluster.

  {\em dw\,1251+2324} A new tiny satellite of nearby spiral galaxy NGC\,4826, which has a radial velocity of $V_h= 409 \pm 1$\,km\,s$^{-1}$ and TRGB distance of $4.41$~Mpc. Their radial velocity difference is $+125 \pm 13$\,km\,s$^{-1}$ and projected separation of $2.15$~deg or $165$~kpc.
  
  {\em SMDG\,1345+3311}  A distant LSB galaxy with a core and a faint halo.

  {\em dw\,1408-0802} A granular dwarf irregular galaxy at the sbf-distance of 3.79$\pm$0.8 Mpc~\citep{Carlsten2026}. The supergalactic coordinates $\text{SGL} = 129.6$~deg and $\text{SGB} = 15.3$ deg place the galaxy at the edge of the Local Void.

  {\em NGC\,5608} This is an isolated Sdm galaxy with a
Tully-Fisher distance of $13\pm4$\,Mpc. It is probably associated with the dwarf irregular galaxy
FASHI\,1420+43 ($V_h = 628$\,km\,s$^{-1}$).

  As shown by radial velocity measurements of the Local Volume candidate galaxies from Table~\ref{tab:lv_galaxies}, only half of them can be classified as Local Volume members.
The most distant galaxies turned out to be those from the \citet{Hunter2025} sample, which had been selected as probable satellites of the nearby luminous galaxies.

  We also measured radial velocities of some nearby dwarf candidates from the recent lists by \citet{Karachentsev2024}, Karachentsev \& Popova (2024) and \citet{Hunter2025} that were not included in the Local Volume database.
These galaxies are presented in Table~\ref{tab:survey_galaxies}.

  {\em Dw\,1059+2836}  This irregular dwarf is probably associated with UGCA\,225 = Mrk\,36 ($V_h = 646$\,km\,s$^{-1}$) and other members of a group around NGC\,3486.

  {\em Dw\,1207+3922, PGC\,4102660, PGC\,4102727, PGC\,4103048, PGC\,4103168} These are dIrr members of the group NGC\,4151 located at a distance of $14.1$~Mpc. Their radial velocity differences and projected separations with respect to NGC\,4151 are: $+74 \pm 15$\,km\,s$^{-1}$ and $137$~kpc, $+60 \pm 15$\,km\,s$^{-1}$ and $494$~kpc, $+40 \pm 10$\,km\,s$^{-1}$ and $91$~kpc, $-120 \pm 10$\,km\,s$^{-1}$ and $398$~kpc, $-53 \pm 10$\,km\,s$^{-1}$ and $92$~kpc, respectively.

  {\em MAGE\,1423+5613, MAGE\,1425+5550} These galaxies were erroneously
assumed by~\citet{Hunter2025} as members of the M\,101 group.

  {\em Dw\,1552-1027} A very isolated galaxy in the Local Void at a kinematic distance of 11.77 Mpc with a rich emission spectrum and a broad H$\alpha$ line (see Fig.~\ref{fig:dw1552_spec}).
  Only two nearby dwarf galaxies are currently known to reside close to the centre of the Local Void: KK\,246 and ALFALFA ZOA\,J1952+1428.
  Both systems are extremely isolated, gas-rich, and dynamically simple, making them key empirical benchmarks for studying dwarf-galaxy evolution under minimal environmental influence~\citep{Rizzi2017}. Their existence demonstrates that star-forming dwarfs can persist deep inside void interiors,
  where galaxy--galaxy interactions, tidal perturbations, and external gas stripping are effectively absent, and where void expansion may dominate the local dynamics.

 \begin{figure}
     \centering
     \includegraphics[width=0.23\textwidth]{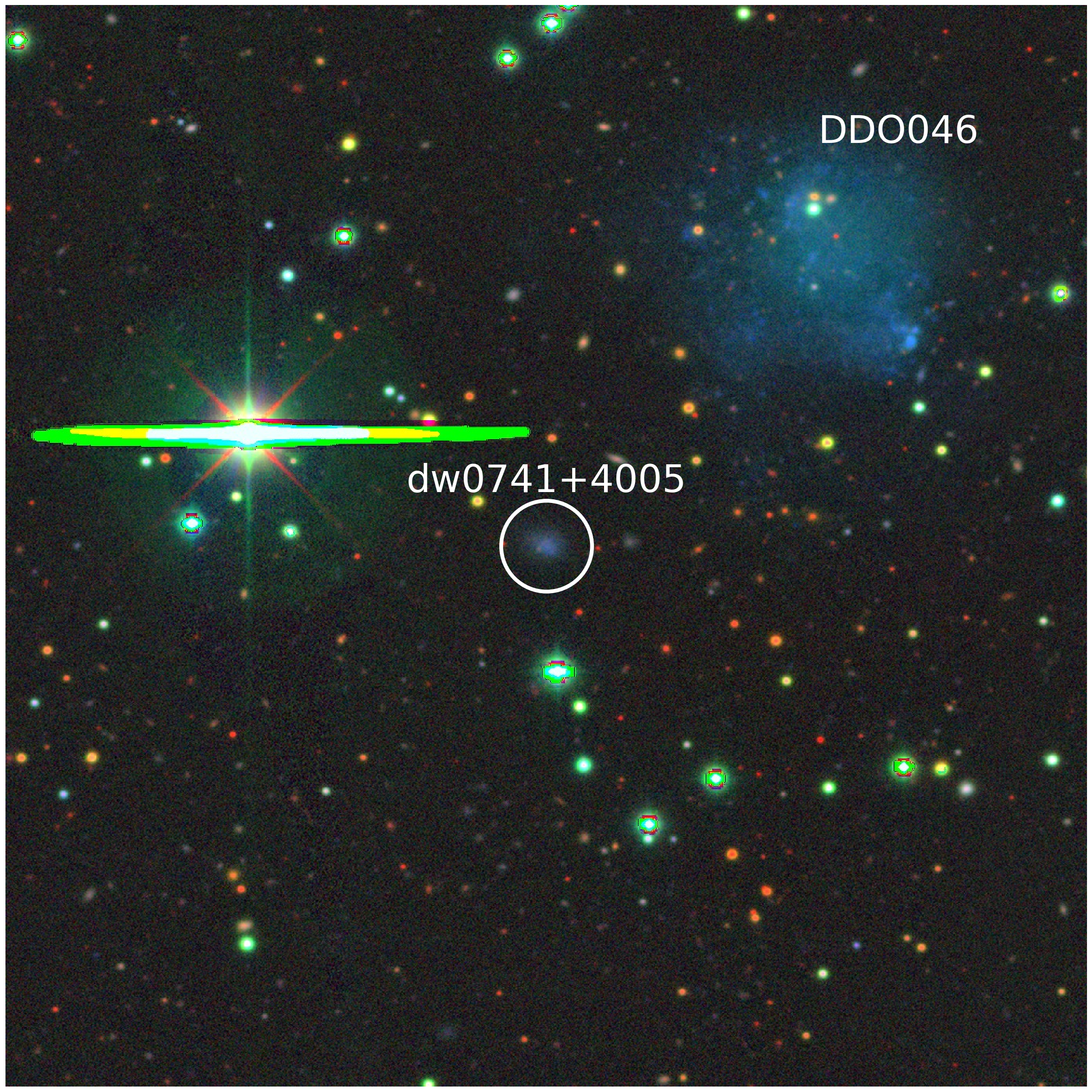}
     \includegraphics[width=0.23\textwidth]{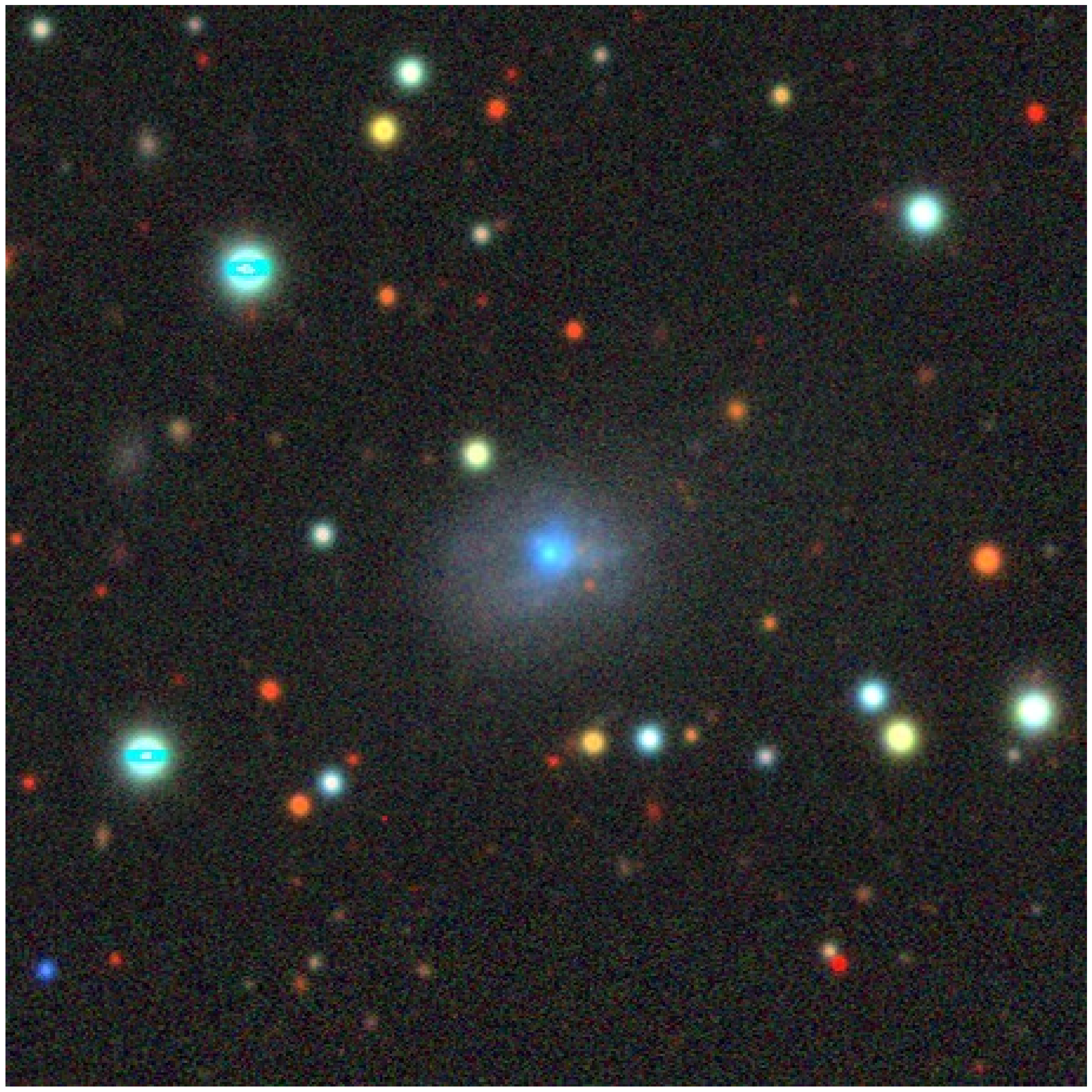}
     \caption{DESI Legacy Survey images of galaxies in this study. Left panel: $6^{\prime}\times6^{\prime}$
         image of DDO\,46 and Dw\,0741+4005; Right panel: $2^{\prime}\times2^{\prime}$ image of Dw\,1552-1027.}
     \label{fig:dw0741+4005}
 \end{figure}

 \begin{figure*}
     \centering
     \includegraphics[width=\textwidth]{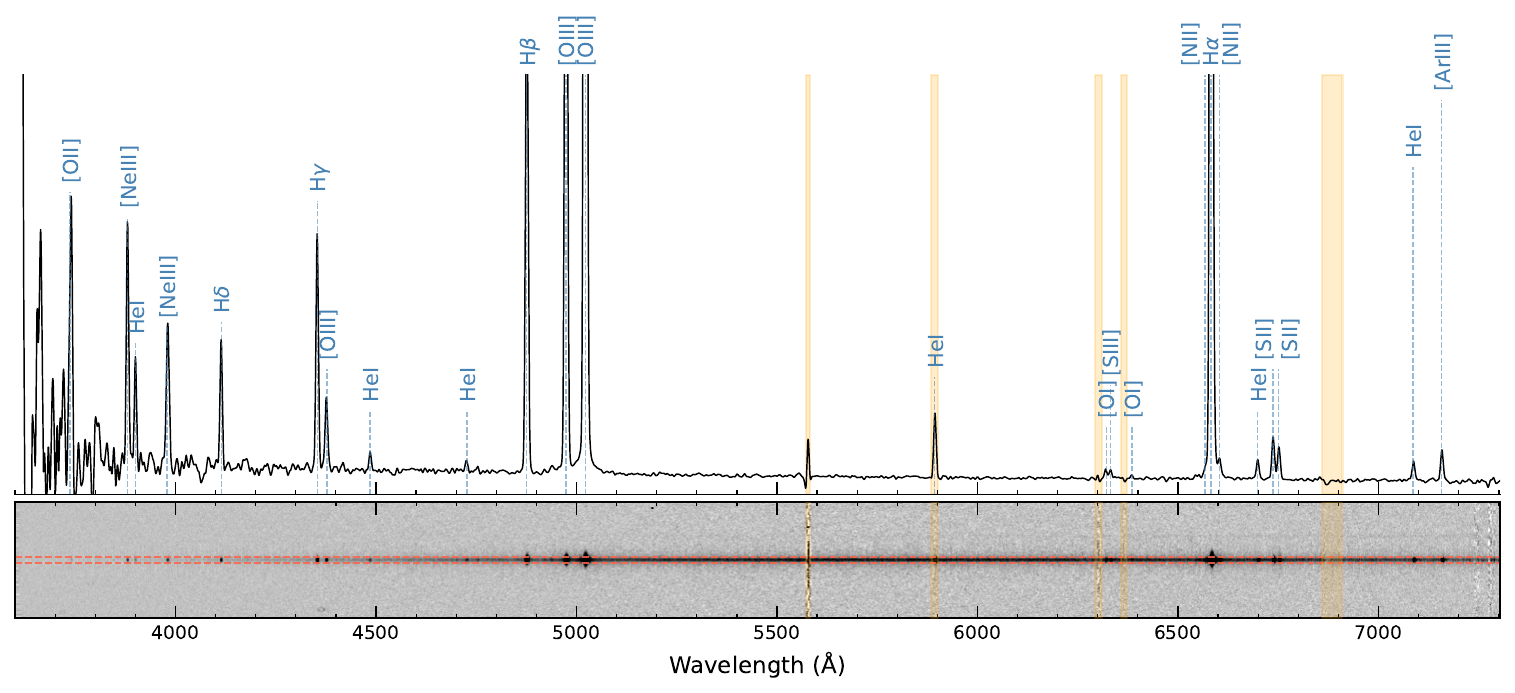}
    \caption{Dw\,1552$-$1027 long-slit spectrum. \textit{Bottom panel:} Two-dimensional spectrum obtained with
    the 6\,m telescope using the SCORPIO-2 spectrograph. The region used for signal extraction is indicated.
    \textit{Upper panel:} Extracted one-dimensional spectrum. Regions with significant residuals from sky-line
    subtraction are highlighted in yellow.}
     \label{fig:dw1552_spec}
 \end{figure*}

\subsection{\texorpdfstring{NGC\,1068 group}{NGC 1068 group}}

The group of galaxies around the luminous spiral galaxy NGC\,1068 was presented
in the catalogue of nearby groups by \citet{Makarov2011}. Based on data for ten of its members with known radial velocities, the authors estimated the radial velocity
dispersion in the group to be $\sigma_v = 80$\,km\,s$^{-1}$, the projected harmonic radius to be 387\,kpc,
the integrated $K$-band luminosity of the group to be $\log(L_K/L_\odot) = 11.54$, and the virial
mass to be $\log(M_v/M_\odot) = 12.79$ at a group distance of 15.9\,Mpc. Similar values for the
parameters of this group were also obtained by \citet{Kourkchi2017}. Over the past
decade, several more members of this group have been discovered due to the ALFALFA
sky survey in the 21\,cm hydrogen line \citep{Haynes2018}.

  We undertook a visual search for new satellites of NGC\,1068 in the sky region bounded by the coordinates $\alpha = [37.0, 44.0]^\circ$ and $\delta = [-3.5, +4.0]^\circ$, using DESI Legacy Imaging Surveys \citep{Dey2019}.
  As a result, we found six new candidate satellites: Dw\,0229$-$0318 $=$ PGC\,1073199, Dw\,0232$-$0256, Dw\,0242$+$0133, Dw\,0242$-$0009, Dw\,0243$-$0108 and Dw\,0244$-$0015.
  The last one was identified as the dwarf spheroidal galaxy SMDG\,0244$-$0015 in the catalogue by \citet{Zaritsky2022}.
Our measurements of radial velocities showed that PGC\,1073199 and Dw\,0232$-$0256 are background galaxies beyond the group, while Dw\,0242$+$0133 and Dw\,0243$-$0108 are true satellites of NGC\,1068.
The dwarf irregular galaxy Dw\,0242$-$0009 has a very low surface brightness,
which makes it difficult to measure its optical velocity.
A list of 17 supposed members of the NGC\,1068 group is presented in Table~\ref{tab:ngc1068_group}.
Three galaxies in the group have individual distance estimates via surface brightness fluctuations~\citep{Carlsten2026}.
Based on the 21\,cm linewidth data, $W$, from LEDA, we estimated the Tully--Fisher distances to eight galaxies using the relation
$M_B = -7.27\log W - 19.99$
\citep{Tully2008},
which is an empirical correlation between a galaxy's rotation velocity -- measured from the width of its 21\,cm line -- and its intrinsic luminosity.

This method yields distance uncertainties of approximately $15 - 20\%$ per spiral disk-dominated galaxy, but for dwarf irregular galaxies these are higher, up to $30 - 50 \%$ depending on their luminosity~\citep{Karachentsev2017}.
For five galaxies, their distances were taken to be the same as the host galaxy's distance derived from Cepheids.

  Distribution of galaxies in the considered sky region is presented in Fig.~\ref{fig:NGC1068}. According to \citet{Markham2026}, the distance of
the group via Cepheids is $10.72\pm0.52$\,Mpc, which is significantly less than that accepted by
\citet{Makarov2011}. The average distance of galaxies in the group derived by TF-
and sbf-methods is $11.48\pm0.55$\,Mpc, which agrees with the Cepheid distance within statistical
errors. However, the average kinematic distance of the group, $\langle D_\mathrm{NAM}\rangle = 13.54\pm0.46$\,Mpc,
is higher.

Based on the Cepheid distance modulus, we obtained the following characteristics for the group:
radial velocity dispersion $\sigma_v = 92$\,km\,s$^{-1}$,
the average projected separation of satellites with known velocities $\langle R_p\rangle = 313$\,kpc,
the average projected estimate of the total mass $\langle M_p\rangle = (2.6\pm1.0)\times10^{12}\,M_{\odot}$.
The projected mass of the host galaxy was determined
according to \citet{Bahcall&Tremaine1981}:
$M_p = \frac{16}{\pi} G^{-1} \langle \Delta V^2 R_p \rangle$,
in which $G$ is the gravitational constant,
$\Delta V$ the difference in radial velocities,
and $R_p$ the projected separations of satellites relative to the host galaxy.
The prefactor on the right-hand side emerges from assuming a random orientation of their orbits
with a mean eccentricity $\langle e^2 \rangle = 1/2$.
With the total luminosity of the group $\log(L_K/L_{\odot}) = 11.25$,
increased by $12\%$ by including the new group members,
and a stellar mass-to-luminosity ratio of $M_*/L_K = 0.6\,M_{\odot}/L_{\odot}$ \citep{Lelli2016},
we obtain a ratio of the total mass of the group to its stellar mass of $\langle M_p/M_*\rangle = 24\pm9$,
typical for other groups in the Local Volume~\citep{Karachentsev2021}.

\begin{figure}
     \centering
     \includegraphics[width=0.45\textwidth]{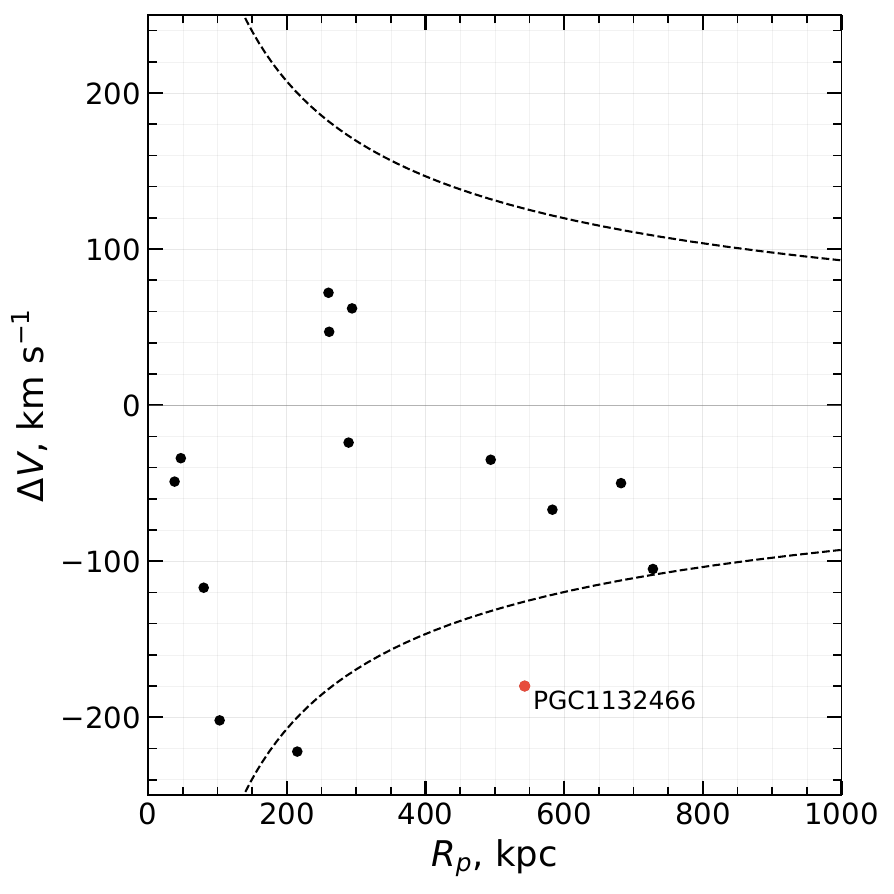}
     \caption{Velocity difference $\Delta V$ relative to NGC\,1068 versus projected radial distance $R_p$ for candidate members of the NGC\,1068 group.
  The dashed curves show the escape velocity for a point mass of $M_p = 10^{12}\,M_\odot$.}
     \label{fig:NGC1068vel}
 \end{figure}

\begin{figure}
     \centering
     \includegraphics[width=0.45\textwidth]{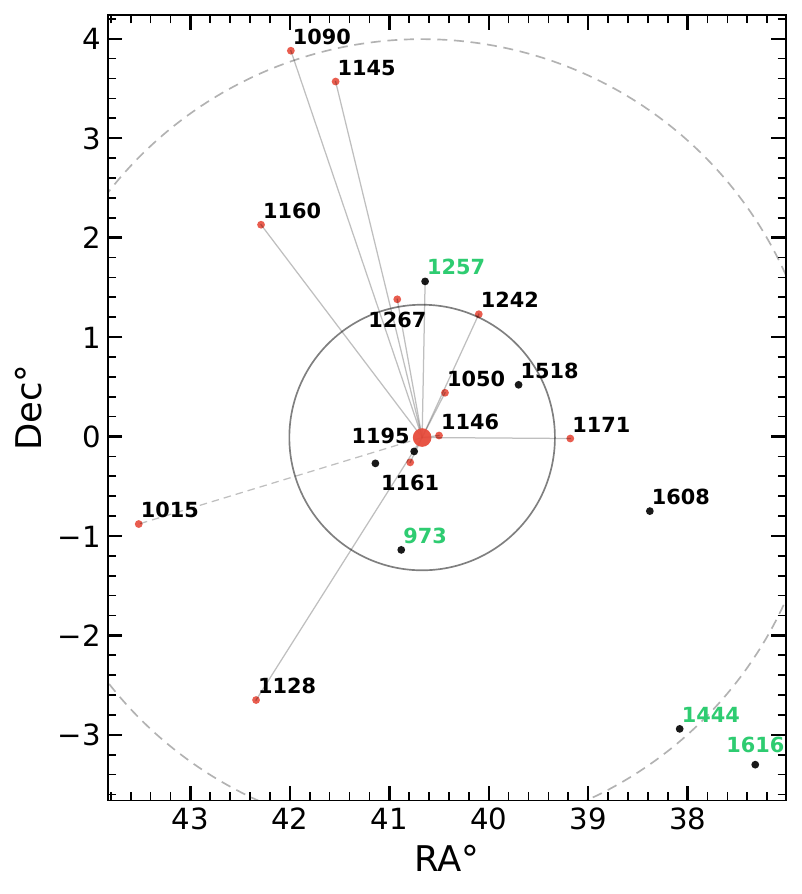}
     \caption{Distribution of galaxies in the NGC\,1068 group.
     Confirmed and candidate members are connected to NGC\,1068 by solid lines;
     PGC\,1132466 is connected with a dashed line as a very unlikely group member;
     newly detected dwarfs are highlighted in green.
     The virial radius $R_v = 296$~kpc is shown by the solid circle, and the circle with three times the virial radius is shown by the dashed line.
     Local Group rest-frame velocities $V_{\mathrm{LG}}$ (km\,s$^{-1}$) are labelled next to each galaxy.}
     \label{fig:NGC1068}
 \end{figure}

Following the recipe by~\citet{Geha2017}~(their Fig. 10), we present in Fig.~\ref{fig:NGC1068vel} the distribution of the NGC\,1068 companions according to radial velocity difference and projected separation relative to the host galaxy. The dashed lines on it correspond to the parabolic velocity for a point-like mass of $1 \times 10^{12}\,M_{\odot}$. As it is seen, one galaxy, PGC\,1132466, at a projected distance of $543$~kpc may be a fictitious member of this group. 
Its inclusion raises the group mass estimate from $(2.6 \pm 1.0) \times 10^{12}\,M_{\odot}$ to $(3.8 \pm 1.5) \times 10^{12}\,M_{\odot}$.

According to \citet{Tully2015}, the virial radius of a group and its total mass are related by an approximate relation
  
\begin{equation}
    (R_v/215\,\mathrm{kpc}) = (M_p/10^{12}\,M_\odot)^{1/3}.
\end{equation}

This yields the value of $R_v = 296$\,kpc or $1.59^\circ$, shown in Fig.~\ref{fig:NGC1068} by the circle. As can be seen, almost all assumed members of
the group are located within ${\sim}3\,R_v$, which corresponds to the radius of the
zero-velocity surface separating the zone of gravitational influence of NGC\,1068 from the
general expanding field.

\begin{figure}
     \centering
     \includegraphics[width=0.45\textwidth]{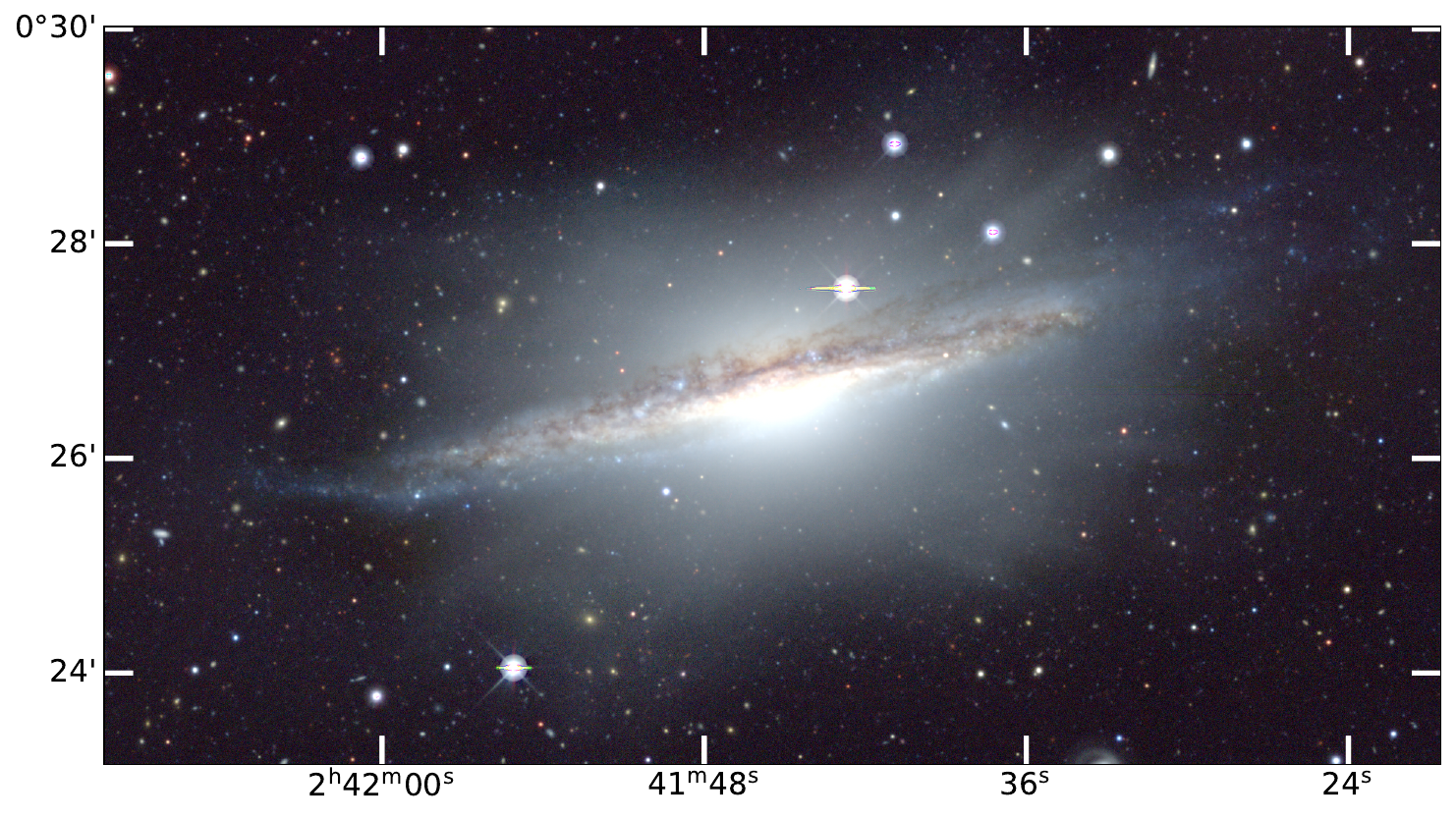}
     \caption{The $12.4\arcmin \times 6.9\arcmin$ optical image of NGC\,1055 obtained from the DESI Legacy Survey Data~\citep{Dey2019}.}
     \label{fig:NGC1055}
 \end{figure}

  It is worth noting that the second-brightest galaxy in the group, NGC\,1055, attracts attention
with the unusual texture of its periphery shown in Fig.~\ref{fig:NGC1055}. Diffuse radial features extending from the centre are
visible at the galaxy's outskirts. The galaxy has an active nucleus and is classified as a Sy\,2 type
galaxy. Note also that among the 17 members of the group, only one dSph satellite with no
signs of current star formation is detected, which is atypical for groups dominated by a luminous
Sb galaxy with a prominent bulge~\citep{Karachentsev2021}.

\section{Conclusions}
\label{sec:conclusions}

Deep wide-field imaging surveys continue to outpace spectroscopic follow-up,
leaving a growing fraction of photometric dwarf candidates without kinematic confirmation.
Our observations demonstrate that long-slit spectroscopy at a large telescope remains an efficient means of closing this gap,
even for low-surface-brightness systems.
We observed 62 dwarf galaxy candidates with the BTA 6-m telescope using the SCORPIO-1 and SCORPIO-2 focal reducers.
Strong emission lines were detected in 34 objects; 12 show only weak H$\alpha$;
16 have no detectable emission.
Radial velocities were measured for 46 galaxies in total.
The main results are as follows.

\begin{itemize}
    \item Of the 26 objects drawn from the Local Volume database (Table~\ref{tab:lv_galaxies}),
    12 have NAM kinematic distances within 13\,Mpc and are new probable companions to the bright Local Volume galaxies.
    The remaining objects from this table turned out to be background galaxies;
    the most distant interlopers are two blue compact galaxies from \citet{Hunter2025} erroneously assumed to be satellites of NGC\,672 and NGC\,784.

    \item Two dIrr galaxies, SMDG\,0740+4032 ($V_h = 363$\,km\,s$^{-1}$) and dw\,0741+4005 ($V_h = 287$\,km\,s$^{-1}$), are new satellites of the nearby SMC-like dwarf DDO\,46 (TRGB distance 10.38\,Mpc).

    \item dw\,0936+2135 is identified as a new dSph satellite of NGC\,2903 (TRGB distance 9.15\,Mpc),
    and dw\,1251+2324 as a new dIrr member of the NGC\,4826 group (TRGB distance 4.41\,Mpc).

    \item Two isolated dIrr galaxies are located at the edge of the Local Void:
    dw\,1408$-$0802 ($V_h = 399$\,km\,s$^{-1}$, $D_\mathrm{sbf} = 3.79$\,Mpc) and Dw\,1552$-$1027 ($V_h = 937$\,km\,s$^{-1}$, $D_\mathrm{NAM} = 11.77$\,Mpc).
    The latter shows a rich emission spectrum with a broad H$\alpha$ component and its radial velocity probably places it in the Local Void.

    \item Five dwarf irregulars --- Dw\,1207+3922, PGC\,4102660, PGC\,4102727, PGC\,4103048, PGC\,4103168 --- are identified as members of the NGC\,4151 group at a distance of 14.1\,Mpc.

    \item A targeted search around NGC\,1068 yielded six new candidate satellites,
    of which Dw\,0242$+$0133 and Dw\,0243$-$0108 are confirmed true group members;
    the group now comprises 15 galaxies with known radial velocities.
    We derive a virial radius $R_v = 296$\,kpc and a projected total mass $\langle M_p\rangle = (2.6\pm1.0)\times10^{12}\,M_\odot$.
    The total-to-stellar mass ratio $\langle M_p/M_*\rangle = 24\pm9$ is consistent with other well-studied Local Volume groups.
\end{itemize}

\section*{Acknowledgements}

This work has made use of DESI Legacy Imaging Surveys data, and the revised version of the Local Volume galaxies database.
The study of LV galaxies has been carried out within the framework of grant of the Russian Science Foundation No 24-12-00277.

Observations with the SAO RAS telescopes are supported by the Ministry of Science and Higher Education of the Russian Federation.
The renovation of telescope equipment is currently provided within the national project "Science and universities".

\section*{Data Availability}

The data underlying this article are publicly available and detailed in the corresponding tables of the manuscript.



\bibliographystyle{mnras}
\bibliography{bta2026} 



\appendix

\section{Tables}

\begin{table*}
    \centering
    \caption{Galaxies from the Local volume database.}
    \label{tab:lv_galaxies}
    \begin{threeparttable}
        \begin{tabular}{lclrrrr}
            \hline\hline

            Name \tnote{1}           & $\textit{RA (2000.0) Dec.}$       & $B$ \tnote{2}  & $V_h$   \tnote{3}    & $V_{\mathrm{LG}}$  \tnote{4} & $V_h^{l}$   & $D_{\mathrm{NAM}}$ \tnote{5} \\

            \hline

            dw\,0139+1433    & \radec{01}{39}{50.6}{+14}{33}{22} & 18.1 & $746\pm10$  & 909           & 757 \tnote{a} , 749.5  \tnote{l}                  & 11       \\
            MAGE\,0148+2830  & \radec{01}{48}{54.0}{+28}{30}{18} & 18.2 & $3677\pm10$ & 3876          & 3677 \tnote{e} , 3687 \tnote{l}                 &             \\
            MAGE\,0202+2717  & \radec{02}{02}{34.8}{+27}{17}{06} & 19.8 & $4795\pm10$ & 4981          & 4808 \tnote{e}                                &             \\
            Dw\,0214+2836    & \radec{02}{14}{09.6}{+28}{36}{47} & 19.2 & $2907\pm10$ & 3088          & 2899 \tnote{a}                                    & 37       \\
            {[TT2009]\,30}   & \radec{02}{22}{54.7}{+42}{42}{45} & 17.5 & $6607\pm10$ & 6818          &                                                         &             \\
            SMDG\,0223-0203    & \radec{02}{23}{18.7}{-02}{03}{25} & 19.6 & $1242\pm15$ & 1305          & 1245 \tnote{a}                                    & 15       \\
            Dw\,0229-0318      & \radec{02}{29}{16.9}{-03}{18}{01} & 16.7 & $1564\pm10$ & 1616          & 1559    \tnote{l}                                       & 19       \\
            Dw\,0232-0256      & \radec{02}{32}{19.7}{-02}{56}{24} & 16.9 & $1393\pm10$ & 1444          & 1409    \tnote{l}                                       & 17       \\
            FASHI\,0237+3855 & \radec{02}{37}{18.7}{+38}{56}{01} & -    & $428\pm10$  & 619           & 420 \tnote{b}                                     & 9        \\
            Dw\,0242+0133      & \radec{02}{42}{33.1}{+01}{33}{50} & 17.3 & $1196\pm10$ & 1256          &                1196 \tnote{a}                                         & 15       \\
            Dw\,0243-0108      & \radec{02}{43}{32.4}{-01}{08}{38} & 17.5 & $925\pm10$  & 973           &                                                         & 12       \\
            PGC\,1166738     & \radec{03}{06}{46.9}{+00}{28}{11} & 18.2 & $706\pm10$  & 739           & 705 \tnote{c} , 689 \tnote{l} , 702 \tnote{s}     & 10       \\
            dw\,0445+0344    & \radec{04}{45}{58.7}{+03}{44}{51} & 18.2 & $692\pm10$  & 647           &                                                         & 12        \\
            SMDG\,0740+4032    & \radec{07}{40}{23.0}{+40}{32}{56} & 19.1 & $363\pm10$  & 374           & 353  \tnote{a}                                    & 11       \\
            MAGE\,0740+1651  & \radec{07}{40}{40.8}{+16}{51}{14} & 20.4 & $4816\pm10$ & 4707          &                                                         &             \\
            dw\,0741+4005    & \radec{07}{41}{34.2}{+40}{05}{03} & 19.0 & $287\pm16$  & 294           &                                                         & 9         \\
            dw\,0936+2135    & \radec{09}{36}{21.4}{+21}{35}{56} & 19.2 & $644\pm38$  & 531 \tnote{*} &                                                         & 13       \\
            dw\,1008+7038b   & \radec{10}{08}{48.1}{+70}{38}{44} & 18.5 & $1301\pm10$ & 1450          &                                                         & 23       \\
            {[KK98]\,108}    & \radec{11}{40}{03.6}{+46}{28}{43} & 18.2 & $770\pm15$  & 811 \tnote{*} &                                                         & 14        \\
            MAGE\,1216+6906  & \radec{12}{16}{15.6}{+69}{06}{25} & 18.2 & $1450\pm10$ & 1608          & 1469 \tnote{l}                                        & 25        \\
            dw\,1237+2602   & \radec{12}{37}{02.8}{+26}{01}{59} & 18.5 & $860\pm10$ & 824 &                                                         &   8          \\
            KDG\,171         & \radec{12}{39}{02.4}{-00}{39}{54} & 17.2 & $1122\pm17$ & 968 \tnote{*} &                                                         &             \\
            dw\,1251+2324    & \radec{12}{51}{28.3}{+23}{24}{03} & 19.1 & $534\pm13$  & 495           &                                                         & 6        \\
            SMDG\,1345+3311    & \radec{13}{45}{11.0}{+33}{11}{31} & 19.7 & $4479\pm13$ & 4520 \tnote{*}&                                                         &             \\
            dw\,1408-0802    & \radec{14}{08}{51.9}{-08}{02}{16} & 16.5 & $399\pm10$  & 284           &                                                         & 3        \\
            NGC\,5608        & \radec{14}{23}{17.9}{+41}{46}{33} & 13.9 & $650\pm10$  & 752           & 663   \tnote{d} , 660 \tnote{l} , 644 \tnote{s}   & 11       \\

            \hline\hline

        \end{tabular}
        \begin{tablenotes}
            \item[1]  The name of the galaxy as it appears in the LVGDB \url{http://www.sao.ru/lv/lvgdb}.
            \item[2]  The apparent $B$-magnitude of the galaxy taken from LEDA or NED, or estimated via $g$ and $r$-magnitudes from DESI
            survey as $B=g+0.313(g-r)+0.227$ according to Lupton\footnote{http://www.sdss3.org/dr10/algorithms/}.
            \item[3] The heliocentric velocity of the galaxy in km~s$^{-1}$.
            \item[4] The radial velocity (km~s$^{-1}$) in the Local Group rest frame.
            \item[5] The kinematic distance of the galaxy in Mpc, determined in the NAM
            \footnote{https://edd.ifa.hawaii.edu/NAMcalculator/} model. The typical uncertainty is about $1.6$~Mpc.
            \item[*] Weak signal.
            \item[] \textit{Literature velocities:} [a]~\citet{Nazarova2025}; [b]~\citet{Karachentsev2024};
            [c]~\citet{vanDriel2016}; [d]~\citet{deVaucouleurs1991}; [e]~\citet{Hunter2026}; [s]~\citet{Ahumada2020}; [l]~\citet{DESICollab2025}. Some of these velocities are 21\,cm measurements.
        \end{tablenotes}
    \end{threeparttable}
\end{table*}

\begin{table*}
    \centering
    \caption{Another candidates to nearby dwarfs from recent sky surveys. All the designations are the same with Table~\ref{tab:lv_galaxies}}
    \label{tab:survey_galaxies}
    \begin{threeparttable}
        \begin{tabular}{lclrrrr}
            \hline\hline
            Name          & $\textit{RA (2000.0) Dec.}$       & $B$  & $V_h$                 & $V_{\mathrm{LG}}$ & $V_h^{l}$                        & $D_{\mathrm{NAM}}$ \\
            \hline

            Dw\,0223+2334   & \radec{02}{23}{16.7}{+23}{34}{32} & 17.9 & $1788\pm10$           & 1946              &                                  & 23   \\
            Dw\,0857+8130     & \radec{08}{57}{00.0}{+81}{30}{00} & 18.3 & $1442\pm10$           & 1641              & 1439 \tnote{l}                   & 23   \\
            Dw\,1059+2836     & \radec{10}{59}{46.2}{+28}{36}{40} & 18.1 & $691\pm10$            & 626               & 706 \tnote{s}                    &         \\
            Dw\,1143+1156     & \radec{11}{43}{56.4}{+11}{56}{24} & 17.1 & $3110\pm25$           & 2980              & 2998    \tnote{s}                &         \\
            Dw\,1207+3922     & \radec{12}{07}{41.3}{+39}{22}{44} & 17.9 & $1071\pm15$           & 1086              & 1079 \tnote{l} , 1067 \tnote{s}  & 19   \\
            PGC\,4102660    & \radec{12}{08}{09.6}{+37}{27}{25} & 18.3 & $1057\pm15$           & 1063              & 1079 \tnote{l} , 1068 \tnote{s}  & 19    \\
            PGC\,4102727    & \radec{12}{08}{41.0}{+39}{19}{23} & 18.1 & $1037\pm10$           & 1052              & 1049 \tnote{l} , 1048 \tnote{s}  & 19    \\
            Dw\,1209+3911   & \radec{12}{09}{59.8}{+39}{11}{46} & 17.4 & $1327\pm10$           & 1342              & 1321 \tnote{s}                   & 25   \\
            PGC\,4103048    & \radec{12}{11}{16.8}{+37}{49}{16} & 16.6 & $877\pm10$            & 886               & 884 \tnote{s}                    & 14   \\
            PGC\,4103168    & \radec{12}{12}{16.8}{+39}{15}{29} & 18.0 & $944\pm10$            & 961               & 953 \tnote{s}                    & 15   \\
            Dw\,1339-1046   & \radec{13}{39}{30.3}{-10}{46}{38} & 16.4 & $1368\pm10$           & 1218              & 1367                             & 18   \\
            MAGE\,1423+5613 & \radec{14}{23}{28.6}{+56}{13}{56} & 20.3 & $1782\pm10$           & 1936              & 1806 \tnote{e}                   & 29   \\
            MAGE\,1425+5550 & \radec{14}{25}{16.8}{+55}{50}{03} & 18.7 & $1824\pm10$           & 1978              & 1790 \tnote{e}                   & 29   \\
            Dw\,1520-1025   & \radec{15}{20}{13.8}{-10}{25}{48} & 18.2 & $2022\pm10$           & 1961              &                                  & 25   \\
            Dw\,1552-1027   & \radec{15}{52}{17.4}{-10}{27}{14} & 17.0 & $937\pm10$            & 905               &                                  & 12   \\
            Dw\,1603-1119   & \radec{16}{03}{19.7}{-11}{19}{35} & 19.5 & $1760\pm17$ \tnote{*} & 1735              &                                  & 22    \\
            Dw\,1721+7140     & \radec{17}{21}{45.4}{+71}{40}{59} & -    & $1086\pm10$           & 1336              &                                  & 19    \\
            Dw\,1810+5238     & \radec{18}{10}{15.8}{+52}{38}{46} & 19.1 & $2884\pm10$           & 3148              &                                  &         \\
            Dw\,2005-1035   & \radec{20}{05}{58.7}{-10}{35}{30} & 18.6 & $1151\pm10$           & 1301              &                                  & 14    \\
            Dw\,2111+0510   & \radec{21}{11}{25.6}{+05}{10}{09} & 17.7 & $3984\pm10$           & 4216              & 3987    \tnote{l}                &         \\

            \hline\hline
        \end{tabular}
    \end{threeparttable}
\end{table*}

\begin{table*}
    \centering
    \caption{Members of the NGC\,1068 group of galaxies. See Table~\ref{tab:lv_galaxies} for column definitions.}
    \label{tab:ngc1068_group}
    \begin{threeparttable}
        \begin{tabular}{lrrcrrrrl@{\hspace{3em}}lr}
            \hline\hline

            Name & $\alpha\,(^\circ)$ & $\delta\,(^\circ)$ & Type & $B$ & $V_h $ & $V_{\mathrm{LG}}$ & $W$ \tnote{1} & $D$ \tnote{2} & $R_p$ \tnote{3} & $M_p/10^{12}$ \tnote{4} \\

            \hline

            AGC\,421098      & 39.18 & $-$00.02 & Im   & 16.47 & $1112\pm  2$ & 1171 &  86 & 13.8\tnote{sbf} &  289 &  0.1 \\
            KDG\,22          & 40.10 & $+$01.23 & Irr  & 16.25 & $1181\pm  2$ & 1242 &  41 &  9.4\tnote{TF}  &  261 &  0.8 \\
            PGC\,5075617     & 40.21 & $+$00.03 & BCD  & 19.0  & $1022\pm 69$ & 1078 &     & 10.7\tnote{mem} &   80 &  1.2 \\
            NGC\,1055        & 40.44 & $+$00.44 & Spec & 11.40 & $ 993\pm  3$ & 1050 & 363 & 14.4\tnote{TF}  &  103 &  2.3 \\
            PGC\,1154903     & 40.50 & $+$00.01 & dE   & 16.65 & $1091\pm  6$ & 1146 &     & 11.6\tnote{sbf} &   38 &  0.1 \\
            Dw\,0242$+$0133  & 40.64 & $+$01.56 & Im   & 17.32 & $1196\pm  2$ & 1256 &  27 & 8.1\tnote{TF}  &  294 &  1.5 \\
            NGC\,1068        & 40.67 & $-$00.01 & Sb   &  9.61 & $1141\pm  6$ & 1195 & 261 & 10.7\tnote{cep} &    0 &      \\
            Dw\,0242$-$0009  & 40.75 & $-$00.15 & Irr  & 19.5  &              &      &     & 10.7\tnote{mem} &   30 &      \\
            PGC\,135659      & 40.79 & $-$00.26 & Im   & 16.85 & $1108\pm  5$ & 1161 &     &  8.9\tnote{sbf} &   47 &  0.1 \\
            Dw\,0243$-$0108  & 40.88 & $-$01.14 & Irr  & 17.5  & $ 925\pm  3$ &  973 &  27 & 11.9\tnote{TF}  &  215 & 12.0 \\
            NGC\,1073        & 40.92 & $+$01.38 & Scd  & 11.47 & $1208\pm  2$ & 1267 &  75 & 10.7\tnote{mem} &  260 &  1.5 \\
            SMDG\,0244$-$0015  & 41.14 & $-$00.27 & dSph & 19.43 &              &      &     & 10.7\tnote{mem} &  100 &      \\
            KDG\,26          & 41.54 & $+$03.57 & Im   & 17.57 & $1080\pm  5$ & 1145 &  29 &  9.5\tnote{TF}  &  682 &  1.7 \\
            DDO\,28          & 41.99 & $+$03.88 & Sdm  & 13.6  & $1025\pm  2$ & 1090 &  85 & 12.0\tnote{TF}  &  728 &  8.8 \\
            DDO\,29          & 42.29 & $+$02.13 & Sm   & 14.1  & $1103\pm  1$ & 1160 &  52 &  9.4\tnote{TF}  &  494 &  0.5 \\
            UGCA\,044        & 42.34 & $-$02.65 & Sm   & 16.9  & $1091\pm  3$ & 1128 &  89 & 13.6\tnote{TF}  &  583 &  2.8 \\
            PGC\,1132466 \tnote{5} & 43.52 & $-$00.88 & BCD  & 17.5  & $ 974\pm  1$ & 1015 &     & 10.7\tnote{mem} &  543 & 20.1 \\

            \hline\hline

        \end{tabular}
        \begin{tablenotes}
            \item[1] FWHM width of the H\,{\sc i} line (km\,s$^{-1}$).
            \item[2] Distance to the galaxy (Mpc). [mem]~assumed membership in the NGC\,1068 group; [cep]~\citet{Markham2026}; [TF]~Tully--Fisher relation ($M_B = -7.27\log W - 19.99$, \citealt{Tully2008}); [sbf]~\citet{Carlsten2026}.
            \item[3] Projected separation from NGC\,1068 (kpc), assuming the same distance as that of the host.
            \item[4] Projected virial mass estimate $M_p = (16/\pi)\,G^{-1}\,\Delta V^2\,R_p$ ($10^{12}\,M_\odot$), where $\Delta V$ is the radial velocity difference relative to NGC\,1068.
            \item[5] This object is probably not the group member.
        \end{tablenotes}
    \end{threeparttable}
\end{table*}

\bsp	
\label{lastpage}
\end{document}